\documentclass[twoside]{elsart}
\usepackage{epsfig}
\usepackage{array}
\usepackage{amssymb,amsbsy,amsmath}
\def\bra#1{\langle \, {#1} \, | \;}
\def\ket#1{\; | \, {#1} \, \rangle}
\newcommand{\braket}[2]{\langle \, {#1} \, | \, {#2} \, \rangle}
\newcommand{\op}[1]{%
    \fontdimen12\textfont3=2pt\fontdimen12\scriptfont3=1.4pt%
    \!\null\mathop{\vphantom{#1}\smash{#1}}\limits_{\sim}\null\!}

\newcommand{\peta}{p_\eta}
\newcommand{\Mean}[1]{\big\langle\big\langle \; {#1}\; 
            \big\rangle\big\rangle}
\newcommand{\SmallMean}[1]{\langle\langle {#1} 
            \rangle\rangle}
\newcommand{\fmref}[1]{(\protect\ref{#1})}
\newcommand{\xref}[1]{\protect\ref{#1}}
\newcommand{\EinsOp}
           {\;\smash{\raisebox{-1.1ex}{$\!\!\stackrel{\!\mbox{1}
            \hspace{-0.4ex}\rule[0.0ex]{0.06ex}{1.60ex}}{\sim}$}}}
\begin{document}
%
\typeout{ --- >>> Nos\'e thermostat for coherent states - paper <<<  --- }
\typeout{ --- >>> Nos\'e thermostat for coherent states - paper <<<  --- }
\typeout{ --- >>> Nos\'e thermostat for coherent states - paper <<<  --- }
%
%
\journal{PHYSICA A}
\begin{frontmatter}
\title{Nos\'e-Hoover dynamics for coherent states}
\author{D. Mentrup\thanksref{DM}}
\author{, J. Schnack}
\address{Universit\"at Osnabr\"uck, Fachbereich Physik \\ 
         Barbarastr. 7, D-49069 Osnabr\"uck}

\thanks[DM]{email: dmentrup\char'100uos.de,\\
            WWW:~http://www.physik.uni-osnabrueck.de/makrosysteme}

\begin{abstract}

\noindent
The popular method of Nos\'e and Hoover to create canonically distributed
positions and momenta in classical molecular dynamics simulations is
generalized to a genuine quantum system of infinite dimensionality. 
We show that for the quantum harmonic oscillator, the equations of
motion in terms of coherent states can easily be modified in an 
analogous manner to mimic the coupling of the system to a thermal bath
and create a quantum canonical ensemble. Possible
applications to more complex systems, especially interacting Fermion
systems, are proposed.

\vspace{1ex}

\noindent{\it PACS:} 
05.30.-d;        
05.30.Ch;        
02.70.Ns         

\vspace{1ex}

\noindent{\it Keywords:} Quantum statistics; Canonical ensemble;
Ergodic behaviour; Thermostat; mixed quantum-classical system
\end{abstract}
\end{frontmatter}
\raggedbottom
\section{Introduction and summary}

The typical problem in statistical physics is the determination of
ensemble averages. The canonical ensemble is characterized
by a constant temperature, i. e. the total energy of the system is
allowed to fluctuate around its mean value, but the system is kept at
a constant temperature by thermal contact with an external heat bath.
Besides the direct evaluation of ensemble averages which is impossible
in many cases, especially in interacting many-body systems, numerous
different approaches have been developed to calculate canonical
ensemble properties, among them Monte Carlo approaches and stochastic
techniques. In classical molecular dynamics, Nos\'e has developed a
scheme that allows to calculate canonical averages by averaging over a
deterministic isothermal time evolution \cite{Nos84,Nos91}. This
scheme is called the classical Nos\'e-thermostat and has attracted
much interest. In the Nos\'e-method, a degree of freedom $s$ and its
conjugate momentum $p_s$ are added to the original system for
temperature control. The additional degree of freedom $s$ acts as a
scaling factor for the positions and momenta of the original system.
The idea is now that the isoenergetic time evolution of the extended
system (that conserves the total energy of the extended system) yields
an isothermal time evolution in the subspace of the variables of the
original system. This holds for ergodic time evolutions. For a
detailed review, see \cite{Nos91}.

Although in practice the original formulation of Nos\'e turned out to
be too cumbersome and featured ergodicity problems in many cases, it
allowed for a number of improvements that led to very effective and
versatile methods \cite{Hoo85}. Simply speaking, the resulting schemes
exploit the equipartition theorem of classical mechanics to determine
the equations of motion of pseudofriction coefficients. The most
reliable methods are the so-called Nos\'e-Hoover chains \cite{Mar92}
and the demon method of Kusnezov, Bulgac, and Bauer (KBB) \cite{KBB90}
for which even the simple one-dimensional harmonic oscillator is
ergodic.

For quantum systems, equivalent methods of comparable power are not
yet available. Grilli and Tosatti have found a theorem that provides a
basis for a seemingly possible translation of the Nos\'e-method to
quantum mechanics \cite{GrT89}. However, in practice, their method
features substantial problems \cite{Mauri93,SMF2001}.  Kusnezov
has proposed a method for finite dimensional quantum systems that can
be applied if all eigenvectors and eigenvalues of the Hamiltonian are
known \cite{Kus93}.

For coherent states, a quantum phase space $(r,p)$ can be defined
properly and a thermal weight function $w_{qm}(\beta;r,p)$ exists that
permits the calculation of canonical ensemble averages as phase space
integrals \cite{Schnack99}. In this article, we present a modification
of the quantum equations of motion of coherent states in a
one-dimensional harmonic oscillator, following closely the ideas of
Nos\'e, Hoover, and KBB for classical systems. In order to calculate
ensemble averages by time averaging, the quantum equations of motion
of the parameters $(r,p)$ of the coherent states are modified. More
precisely, a classical pseudofriction coefficient $\peta$ is added to
the system and the equations of motion are designed in such a way that
the distribution
\begin{eqnarray}
f(\beta;r,p,\peta) \propto 
w_{qm}(\beta;r,p) \; \exp \left
( - \beta \frac{\peta^2}{2Q} \right) 
\nonumber
\end{eqnarray}
which is defined on the mixed quantum-classical phase space
$(r,p,\peta)$ is a stationary solution of a generalized Liouville
equation. As a consequence, if the system is ergodic, $f$ is the
stationary probability distribution generated by the modified quantum
dynamics, and canonical ensemble averages can be calculated by time
averages over the trajectories thus generated. Hence, our method
provides an isothermal quantum time evolution for a quantum system of
infinite dimensionality. It is straightforward to generalize it to
systems of many distinguishable particles in a three-dimensional
harmonic oscillator as well as to free particles. Even non-interacting
fermions may be thermalized, since in this case, the quantum
distribution function is also known \cite{PhDSchnack}.


\section{Method and setup}

\subsection{Coherent states in a harmonic oscillator potential} 
Given the Hamilton operator $\op{H}$ of the one-dimensional harmonic
oscillator
\begin{eqnarray}
\label{E-1}
\op{H} =
\hbar\omega \; 
\left( \op{a}^\dagger \op{a} + \frac{1}{2} \right)
\ ,
\end{eqnarray}
coherent states are defined as eigenstates of the destruction
operator $\op{{a}}$
\begin{eqnarray}
\label{E-2}
\op{a}\; \ket{z} =  z \; \ket{{z}}
\quad , \qquad
z = 
\sqrt{\frac{m \omega}{2 \hbar}} \; r
+ \frac{i}{\sqrt{2 m \hbar \omega}} \; p
\ .
\end{eqnarray}
A coherent state is labelled by its complex eigenvalue $z$ which
corresponds to a pair of real parameters $(r,p)$. Explicitly, in
coordinate representation, coherent states are shifted Gaussian wave
packets characterized by the parameters $r$ (mean position) and $p$
(mean momentum):
\begin{eqnarray}
\label{E-3}
\braket{x}{z}=
\braket{x}{r,p}
\propto
\exp \left\{ -\frac{(\,x-r\,)^2}{2} \frac{m \omega}{\hbar}
+\frac{i}{\hbar} p  \, x \right\} 
\ .
\end{eqnarray}
Coherent states have been extensively investigated \cite{Klauder}. In
particular, the following equality is useful for considering the time
evolution of coherent states in a harmonic oscillator potential:
\begin{eqnarray}
\label{E-4}
\exp(-i \, \omega \, \op{a}^\dagger \op{a} \, t) \ket{z} 
= \ket{\exp(-i \omega t) z}
\ .
\end{eqnarray}
This implies that the exact quantum time evolution of a coherent state
in a harmonic oscillator potential is given by the following equations
of motion for the parameters $r$ and $p$
\begin{eqnarray}
\label{E-5}
\frac{d}{dt}r = \frac{p}{m} \quad , \qquad
\frac{d}{dt}p = -m \omega^2 r 
\ .
\end{eqnarray}
We stress that in a harmonic oscillator potential the solution of
these two coupled ordinary differential equations provides the {\it
  exact quantum time evolution of coherent states}.

Furthermore, the set of all coherent states forms an overcomplete
basis of the Hilbert space with $\int
\frac{\mbox{\scriptsize{d}}r\,\mbox{\scriptsize{d}}p}{(2\pi\hbar)}
\;\ket{r,p}\bra{r,p} = \EinsOp$. As a consequence, given an observable
$\op{B}$, its thermodynamic mean value may be evaluated using coherent
states:
\begin{eqnarray}
\label{E-6}
\Mean{\op{B}}
&=& 
\frac{1}{Z(\beta)} \;
\mbox{tr} \left( \op{B} e^{-\beta \op{H}} \right) 
\\
&=&
\frac{1}{Z(\beta)}
\int  \frac{\mbox{d}r\,\mbox{d}p}{(2\pi\hbar)}\;
\bra{r,p}\op{B} e^{-\beta\op{H}} \ket{r,p} 
\nonumber
\ ,
\end{eqnarray}
where $\beta = \frac{1}{k_B T}$ is the inverse temperature,
$Z(\beta)=\mbox{tr} \left( e^{-\beta \op{H}} \right)$ is the usual
canonical partition function and $\SmallMean{\cdot}$ denotes canonical
averages.

As shown in \cite{Schnack99}, one can interpret the space of the
continuous parameters $r$ and $p$ as a phase space and rewrite
\fmref{E-6} as a phase space integral with the thermal weight function
$w_{qm}(\beta;r,p)$
\begin{eqnarray}
\label{E-7}
\Mean{\op{B}}
&=& 
\frac{1}{Z(\beta)}
\int \frac{\mbox{d}r\,\mbox{d}p}{(2\pi\hbar)}\;
w_{qm}(\beta;r,p) \; {\mathcal B}(r,p) \ ,
\\
\label{E-8}
Z(\beta)
&=& 
\int \frac{\mbox{d}r\,\mbox{d}p}{(2\pi\hbar)}\;
w_{qm}(\beta;r,p)\;
\ ,
\end{eqnarray}
with
\begin{eqnarray}
\label{E-9}
{\mathcal B}(r,p)
&=&
\bra{r,p}\op{B} \ket{r,p} \ , 
\\
\label{E-10}
w_{qm}(\beta;r,p)
&=&
e^{-|z|^2
\left(e^{\beta\hbar\omega}-1\right)}
=
e^{-(\frac{p^2}{2m}+\frac{1}{2}m \omega^2 r^2)
\left( e^{\beta\hbar\omega}-1\right) / (\hbar\omega)}
\ .
\end{eqnarray}
The function $w_{qm}$ contains all quantum statistical properties of
the system. From \fmref{E-10} it can be inferred that formally, it
differs from the classical distribution function of the harmonic
oscillator by the factor
$\left(e^{\beta\hbar\omega}-1\right)/(\beta\hbar\omega)$. Note that
this factor tends to $1$ in both the classical ($\hbar \to 0$) and the
high-temperature ($\beta \to 0$) limit.

\subsection{Modification of the equations of motion}
The idea of our method is to modify the equations of motion
\fmref{E-5} of the coherent states in such a way that the distribution
function $w_{qm}$ is sampled provided the time evolution is ergodic.
To this end, we proceed in a way which is in close analogy to the
approaches in classical molecular dynamics \cite{Hoo85,Mar92,KBB90}.
The equation of motion of the parameter $p$ is supplemented by a term
similar to a frictional force. The time evolution of the
pseudofriction coefficient is then determined by the condition that
the desired distribution function is a stationary solution of a
generalized Liouville equation in the generalized phase space.

\subsubsection{Nos\'e-Hoover thermostat and Nos\'e-Hoover chain}
Adopting the notation of Martyna et al. \cite{Mar92}, we investigate
the following analogue of the classical Nos\'e-Hoover dynamics for the
quantum dynamics of coherent states:
\begin{eqnarray}
\label{E-11}
\frac{d}{dt}r = \frac{p}{m} \quad , \qquad
\frac{d}{dt}p = -m \omega^2 r - p \frac{\peta}{Q} 
\ .
\end{eqnarray}
The key point is the equation of motion of the pseudofriction
coefficient $\peta$. It is determined by the condition that the
distribution function
\begin{eqnarray}
\label{E-12}
f(\beta;r,p,\peta) &\propto& 
w_{qm}(\beta;r,p) \; \exp(-\beta \frac{\peta^2}{2Q})
\\
&\propto&
\exp \left( -( \frac{p^2}{2m}+\frac{1}{2}m \omega^2 r^2 )
  \frac{e^{\beta \hbar \omega}-1}{\hbar \omega} - \beta
  \frac{\peta^2}{2Q} \right)
\nonumber
\end{eqnarray}
is a stationary solution of the following generalized Liouville
equation in the mixed quantum-classical phase space
$\Gamma=(r,p,\peta)$:
\begin{eqnarray}
\label{E-13}
\frac{d}{dt}f 
&=&
- f \cdot \left( \frac{\partial}{\partial \Gamma} 
\cdot \dot{\Gamma} \right)
\\
&=&
- f \cdot \left( \frac{\partial}{\partial r} \dot{r} 
+ \frac{\partial}{\partial p}
\dot{p} + \frac{\partial}{\partial \peta} \dot{\peta} \right) 
\ .
\nonumber
\end{eqnarray}
We calculate the left-hand side of \fmref{E-13}, employing the
equations of motion of $r$ and $p$, \fmref{E-11}:
\begin{eqnarray}
\label{E-14}
\frac{d}{dt}f 
&=& \frac{\partial f}{\partial p} \dot{p}
+ \frac{\partial f}{\partial r} \dot{r}
+ \frac{\partial f}{\partial \peta} \dot{p}_\eta 
\\
&=& f \cdot \left( - \left( \frac{p}{m} \dot{p} + m \omega^2 r \dot{r}\right) 
\frac{e^{\beta \hbar \omega}-1}{\hbar \omega}
-\beta\frac{\peta}{Q} \dot{p}_\eta
\right) 
\nonumber \\
&=& f \cdot \left( \frac{p^2}{m} \frac{\peta}{Q} \frac{e^{\beta
    \hbar \omega}-1}{\hbar \omega}-\beta \frac{\peta}{Q}\dot{\peta}
\right)
\ .
\nonumber
\end{eqnarray}
On the right-hand side of \fmref{E-13}, we have the freedom to impose
the constraint $\partial \dot{\peta}/ \partial \peta =0$ that is
common in this context \cite{Nos91}. We obtain
\begin{eqnarray}
\label{E-15}
-f \cdot \left(\frac{\partial}{\partial \Gamma} 
\cdot \dot{\Gamma} \right)
=
f \; \frac{\peta}{Q}
\ .
\end{eqnarray}
Equating \fmref{E-14} and \fmref{E-15} yields the following equation
of motion for $\peta$:
\begin{eqnarray}
\label{E-16}
\frac{d}{dt}\peta &=& \frac{1}{\beta} \left( \frac{p^2}{m}
\frac{e^{\beta \hbar \omega}-1}{\hbar \omega}-1 \right)
\ .
\end{eqnarray}
Again, the only difference between this equation and its classical
counterpart is given by the factor $(e^{\beta \hbar \omega}-1)/(\beta
\hbar \omega)$. Moreover, \fmref{E-16} retains the property of its
classical counterpart that the time evolution of the pseudofriction
coefficient is governed by the deviation of the actual value of a
quantity related to the kinetic energy from its canonical average
value. This can be inferred by evaluating
\begin{eqnarray}
\label{E-17}
\Mean{\frac{p^2}{m}
\frac{e^{\beta \hbar \omega}-1}{\hbar \omega}}
   = 1
\end{eqnarray}
using \fmref{E-7}.

Finally, it is easily confirmed that the set of dynamical equations
\fmref{E-11}, \fmref{E-16} conserve the quantity
\begin{eqnarray}
\label{E-18}
H^\ast = (\frac{p^2}{2m}+\frac{1}{2}m \omega^2 r^2)
\frac{e^{\beta \hbar \omega}-1}{\beta \hbar \omega}
+\frac{\peta^2}{2Q} 
+\int^t \mbox{d}t' \frac{\peta(t')}{\beta}
\ .
\end{eqnarray}
The equations \fmref{E-11} and \fmref{E-16} form a genuine quantum
Nos\'e-Hoover thermostat for coherent states. Since in classical
molecular dynamics, these equations of motion frequently feature
ergodicity problems, Martyna et al. have developed the idea of a chain
thermostat \cite{Mar92}. This method implies to impose another
thermostating pseudofriction coefficient on $\peta$ which may be
coupled to yet another pseudofriction coefficient, and so on, thereby
forming a chain of thermostats. The application of the idea to the
quantum case does not infer anything new compared to the classical
case, since only the first pseudofriction coefficient of the chain
interacts with the quantum phase space variables. Therefore, for
further particulars we refer the reader to \cite{Mar92}.

\subsubsection{KBB-method}
Another generalization of the Nos\'e-Hoover thermostat that is
frequently used in classical molecular dynamics is the so-called demon
method proposed by Kusnezov, Bulgac, and Bauer \cite{KBB90}. The
advantage of this method is that the Hamilton function of the
envisaged system does not have to contain a kinetic energy term for
temperature control; instead, the time derivative of the temperature
control variables is postulated to be proportional to the difference
of two arbitrary quantities whose ratio of canonical averages is
$1/\beta$.

At least two pseudofriction coefficients, so-called demons, are
introduced for temperature control. Both the equations of motion for
positions and momenta are supplemented by additional terms. We
introduce the demons into the quantum equations of motion of the
parameters of coherent states:
\begin{eqnarray}
\label{E-19}
\frac{d}{dt}r = \frac{p}{m} -g_2'(\xi) F(r,p) \quad , \qquad 
\frac{d}{dt}p = -m \omega^2 r - g_1'(\zeta) G(r,p)
\ .
\end{eqnarray}
$F(r,p), G(r,p)$ are arbitrary functions of the quantum phase space
variables. $g_1(\zeta), g_2(\xi)$ are functions of the demons which
have to be chosen so that the integration of the distribution function
$f$ converges. $g_1',g_2'$ are the respective derivatives. The
distribution function on the phase space $\Gamma=(r,p,\zeta,\xi)$
reads
\begin{eqnarray}
\label{E-20}
f(r,p,\xi,\zeta) = 
\exp \left(
-(\frac{p^2}{2m}+\frac{1}{2}m \omega^2 r^2 )
\frac{e^{\beta \hbar \omega}-1}{\hbar \omega}
-\beta 
(\frac{g_2(\xi)}{\kappa_2}+\frac{g_1(\zeta)}{\kappa_1})
\right) 
\end{eqnarray}
and the time evolution of the demons is, as above, deduced from the
requirement that $f$ is a solution of a generalized Liouville equation
in the phase space. We obtain
\begin{eqnarray}
\label{E-21}
\frac{d}{dt} \zeta 
&=&
\kappa_1 \left( \frac{p}{m}G \frac{e^{\beta \hbar
    \omega}-1}{\beta \hbar \omega}-\frac{1}{\beta} \frac{\partial
  G}{\partial p} \right) 
\ ,\\
\label{E-22}
\frac{d}{dt} \xi 
&=&
\kappa_2 \left( m \omega^2 r F \frac{e^{\beta \hbar
    \omega}-1}{\beta \hbar \omega}-\frac{1}{\beta} \frac{\partial
  F}{\partial r} \right)
\ .
\end{eqnarray}
Again, it is interesting to notice that
\begin{eqnarray}
\label{E-23}
\frac{1}{\beta} \Mean{\frac{\partial G}{\partial p}} 
= \Mean{\frac{p}{m}G \frac{e^{\beta \hbar
    \omega}-1}{\beta \hbar \omega}}
\ ,
\end{eqnarray}
i.e. the ratio of the canonical averages of the quantities that
determine the time derivative of the demons is $\beta$, just as in the
classical case. The quantity
\begin{eqnarray}
\label{E-24}
H^\ast &=& (\frac{p^2}{2m}+\frac{1}{2}m \omega^2 r^2)
\frac{e^{\beta \hbar \omega}-1}{\beta \hbar \omega}-
\frac{g_2(\xi)}{\kappa_1}-\frac{g_1(\zeta)}{\kappa_2} \\
&\phantom{=}&
+\frac{1}{\beta} \int^t \mbox{d}t' 
\left( \frac{\partial G}{\partial p}g_1'
+\frac{\partial F}{\partial r}g_2'
\right) 
\nonumber
\end{eqnarray}
is conserved during the time evolution defined by \fmref{E-19},
\fmref{E-21}, \fmref{E-22}.

In principle, since the choice of the functions $F,G,g_1,g_2$ is
arbitrary, this method offers a lot of freedom. The most prominent
coupling scheme recommended by KBB is the so-called cubic coupling
scheme with the following choice of functions \cite{KBB90}
\begin{eqnarray}
\label{E-25}
g_1 = \frac{1}{2} \xi^2 \quad , \qquad
g_2 = \frac{1}{4} \zeta^4 \quad , \qquad 
F = r^3 \quad , \qquad
G = p \quad ,
\end{eqnarray}
which leads in the quantum case to the special set of equations of
motion
\begin{eqnarray}
\label{E-26}
\frac{d}{dt}r &=& \frac{p}{m} -\xi r^3 \quad , \qquad
\frac{d}{dt}p = -m \omega^2 r - \zeta^3 p \\
\label{E-27}
\frac{d}{dt} \zeta &=&
\kappa_1 \left(\frac{p^2}{m} \frac{e^{\beta \hbar
    \omega}-1}{\beta \hbar \omega}-\frac{1}{\beta} \right) \\
\label{E-28}
\frac{d}{dt} \xi &=&
\kappa_2 \left( m \omega^2 r^4 \frac{e^{\beta \hbar
    \omega}-1}{\beta \hbar \omega}-\frac{1}{\beta} 3r^2 \right)
\end{eqnarray}
that we have investigated taking $\kappa_1=\kappa_2=1$.  Finally, we
note that \fmref{E-27}, \fmref{E-28} may easily be linked to the
equations of motion proposed by Kusnezov in \cite{Kus93}.
$w_{qm}$ plays the role of Kusnezov's $\rho(Q,P)$. However, while
Kusnezov's approach is limited to quantum systems of finite
dimensionality, our method works for this system with a Hilbert space
of infinite dimensionality because we take advantage of the properties
of coherent states.


\section{Results}

In classical molecular dynamics simulations of the harmonic
oscillator, the simple Nos\'e-Hoover method features ergodicity
problems, while the Nos\'e-Hoover chain method and the demon approach
of KBB work well \cite{Mar92,KBB90}. The correct classical phase space
density is perfectly reproduced both by the chain and by the demon
dynamics.

Formally, the only difference between the quantum phase space density
$w_{qm}$, eq. \fmref{E-10}, and its classical counterpart is given by the
factor $\left(e^{\beta\hbar\omega}-1\right)/(\beta\hbar\omega)$. This
also applies to the respective equations of motion of the different
dynamics.  Since this factor is only a number that depends on
temperature, but not on the phase space variables $r$ and $p$, we
anticipate that it does not influence the overall characteristics of
the dynamics.  Therefore, we expect that ergodicity problems in the
quantum case will arise under the same circumstances as in the
classical case.

We show results for the set of parameters chosen in \cite{Mar92} to
enable a direct comparison.  We took $m=1, \omega=1$ and initial
conditions $r(0)=1, p(0)=1$ with $1/\beta=1.0$. The numerical
integration of the equations of motion was carried out with a
fourth-order Runge-Kutte algorithm with a step size that ensured
conservation of pseudoenergy to more than seven significant figures.
All runs were made over a total integration time of $2000 \tau$, where
$\tau=2 \pi/\omega$.

\subsection{Nos\'e-Hoover and Nos\'e-Hoover chain method, \newline
 KBB method with cubic coupling scheme}
\begin{figure}
\begin{center}
\epsfig{file=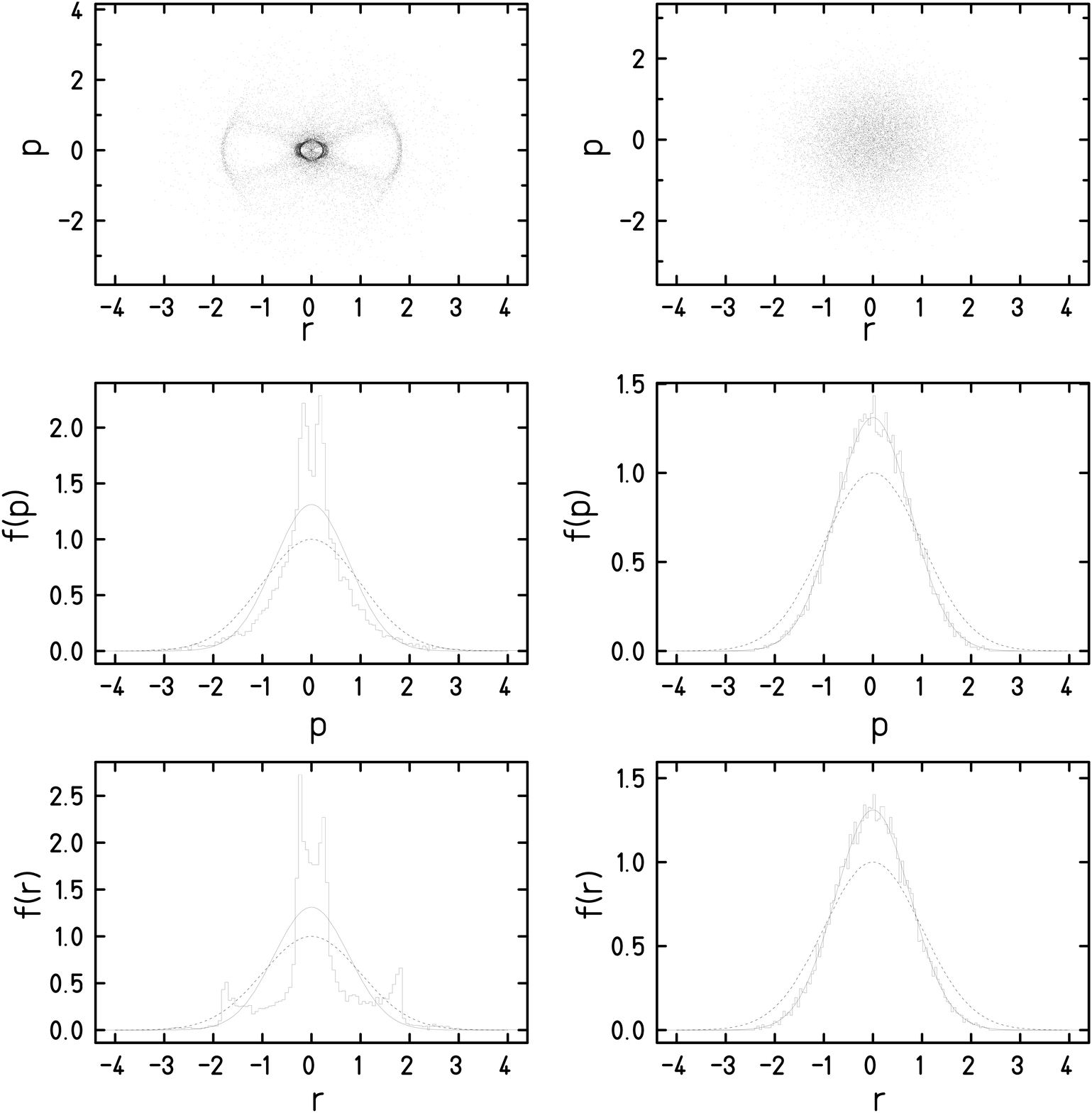,width=130mm}
\caption{Left panel: Simple Nos\'e-Hoover dynamics of a quantum
  harmonic oscillator, right panel: Nos\'e-Hoover chain dynamics. From
  above: $(r,p)$-density plot, momentum distribution, position
  distribution. The solid line depicts the exact quantum result given
  by the respective partially integrated function $w_{qm}/Z$ (e.g.,
  $f(r)=\frac{1}{Z} \int \frac{\mbox{\scriptsize{d}} p}{\sqrt{2 \pi
      \hbar}} w_{qm}(r,p)$), the dashed line represents the
  corresponding classical distribution $w_{cl}$ normalized to the same
  value (see \fmref{E-29}). The distributions sampled by time
  averaging are presented as histograms.}
\label{F-1}
\end{center}
\end{figure}
\begin{figure}
\begin{center}
\epsfig{file=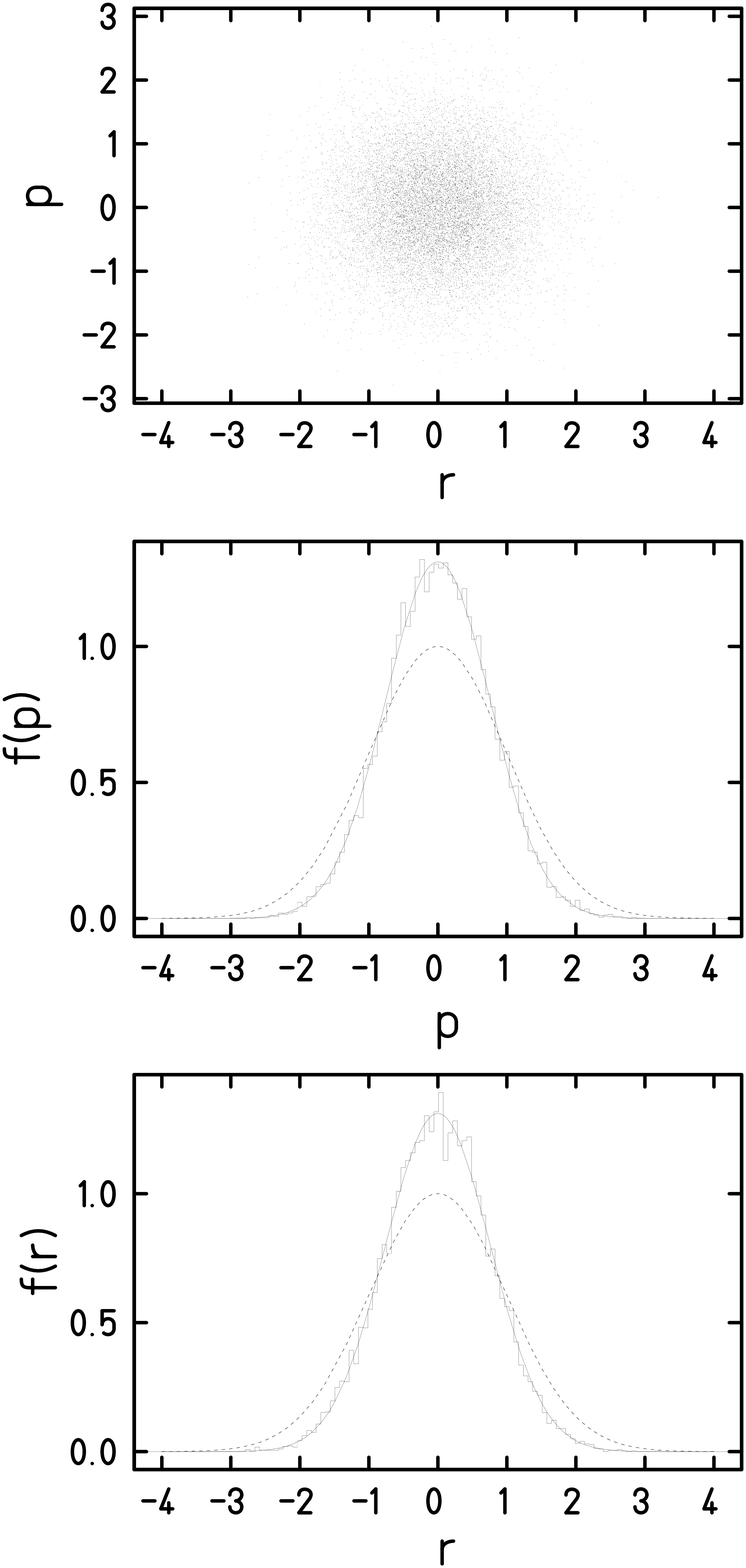,width=65mm}
\caption{KBB dynamics of a quantum harmonic oscillator, description as
  in figure \xref{F-1}.} 
\label{F-2}
\end{center}
\end{figure}
The left panel of figure \xref{F-1} presents a $(r,p)$-density map and
the projected distribution functions for the simple Nos\'e-Hoover
dynamics. We find a result that is very similar to the classical case
\cite{Mar92}: The dynamics does not fill the phase space with the
correct weight and, moreover, we find that the obtained distributions
strongly depend on the initial conditions and values of the parameters
chosen (not shown). Thus, the dynamics is not ergodic.

The situation changes radically with the introduction of a second
thermostating variable acting upon the first pseudofriction
coefficient $\peta$, see figure \xref{F-1}, right panel. The
distribution functions sampled by this time evolution reproduce
$w_{qm}$ extremely well, and changes of the initial conditions and
parameters do not have a noticeable effect on the results. The
dynamics generated in this way is obviously ergodic.  The addition of
further thermostating variables does not influence the results.

We point out that the statistics obtained by time averaging over the
quantum time evolution is the quantum statistics of the harmonic
oscillator. To make this evident, we present plots of the partially
integrated distribution function $w_{qm}(r,p)/Z$ along with plots of
its classical limit
\begin{eqnarray}
\label{E-29}
w_{cl}(r,p) =
\lim_{\frac{e^{\beta\hbar\omega}-1}{\beta\hbar\omega} \to 1}
\frac{1}{Z(\beta)} \, w_{qm}(r,p) =
\beta \hbar \omega 
\exp \left( 
-\beta ( \frac{p^2}{2m} + \frac{1}{2} m \omega^2 r^2 ) 
\right)
\end{eqnarray}
which is proportional to the classical canonical ensemble distribution
function. Since $\left(e^{\beta\hbar\omega}-1\right) /
(\beta\hbar\omega) > 1$ for all $\beta$, the quantum distribution
function is always narrower compared to its classical limit.

Figure \xref{F-2} presents the results obtained from a
KBB-demon-dynamics using the cubic coupling scheme. The results are
similar to the case of the chain dynamics, in particular, the dynamics
is also ergodic.

\subsection{Mean values of selected observables}
\begin{figure}
\begin{center}
\epsfig{file=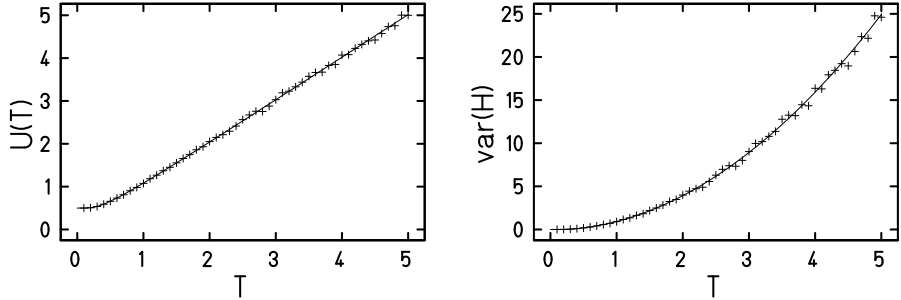,width=120mm}
\caption{Values of the internal energy and its variance of the
  harmonic oscillator obtained from time averaging with a KBB dynamics
  (crosses) compared to the exact quantum canonical ensemble result
  (solid line).}
\label{F-3}
\end{center}
\end{figure}

Finally, the results of time averaging are compared to the analytical
ensemble averages for two typical observables, the internal energy and
its variance. The analytical formulas are briefly given:
\begin{eqnarray}
\label{E-30}
U(\beta) &=& \Mean{\op{H}} = 
\frac{\hbar \omega}{2}
+ \frac{\hbar \omega}{e^{\beta \hbar \omega}-1} 
\\
\mbox{var}(\op{H})&=&
\Mean{\op{H}^2}-\Mean{\op{H}}^2 = 
\left( 
\frac{\hbar \omega}{2 \sinh(\frac{1}{2} \beta \hbar \omega)} 
\right)^2
\nonumber
\end{eqnarray}
Figure \xref{F-3} shows the excellent agreement between the exact
results and the results obtained by time averaging with a KBB
dynamics. The small deviations are clearly of statistical origin. They
increase at higher temperatures because we kept the total sampling
time constant, although the volume of the relevant phase space
increases with temperature. Therefore, to cover it with the same
accuracy at a higher temperature, one would need a longer sampling
time.


\section{Discussion and outlook} 

This article presents a straightforward, yet non-trivial extension of
the powerful methods of heat bath coupling in classical molecular
dynamics simulations to a genuine quantum system of infinite
dimensionality. The application of the method to a quantum system of
many distinguishable particles or a three-dimensional harmonic
potential is a simple generalization.

Since the knowledge of $w_{qm}$ is indispensable for the setting up of
the equations of motion for the pseudofriction coefficients, the
method is limited to systems where $w_{qm}$ is known. Therefore, also
non-interacting identical fermions, both moving freely or contained in
a harmonic oscillator potential, can be thermalized using the
respective distribution functions \cite{PhDSchnack}. Moreover, by
coupling one of the solvable systems to a more complex system of
interacting particles, one can possibly determine its equilibrium
properties.  This idea that permits to evaluate ensemble averages by
time averaging is potentially very powerful, since efficient
approximate quantum dynamics methods (Time-Dependent Hartree-Fock,
Fermionic Molecular Dynamics, etc. \cite{RMP}) are available that are
applicable also for indistinguishable fermions. Thus, a new method of
calculating thermodynamic properties of interacting Fermion systems
seems conceivable that might also work where other methods fail.


{\bf Acknowledgments}\\[5mm] The authors would like to thank the
Deutsche Forschungsgemeinschaft (DFG) for financial support of the
project ``Isothermal dynamics of small quantum systems". We are
thankful to K.~B\"arwinkel for carefully reading the manuscript.


\end{document}